\def\year{2021}\relax
\documentclass[letterpaper]{article} 
\usepackage{aaai21}  
\usepackage{times}  
\usepackage{helvet} 
\usepackage{courier}  
\usepackage[hyphens]{url}  
\usepackage{graphicx} 
\usepackage{natbib}  
\usepackage{caption} 
\usepackage{makecell}
\usepackage{subfigure}
\usepackage{graphicx}
\usepackage{multirow}
\usepackage{amsmath}
\usepackage{amssymb}
\usepackage{booktabs}
\title{FairRec: Fairness-aware News Recommendation with\\ Decomposed Adversarial Learning}

\author{
Chuhan Wu\textsuperscript{\rm 1}, 
Fangzhao Wu\textsuperscript{\rm 2}, 
Xiting Wang\textsuperscript{\rm 2}, 
Yongfeng Huang\textsuperscript{\rm 1},
Xing Xie\textsuperscript{\rm 2} \\
}
\affiliations{
\textsuperscript{\rm 1}Department of Electronic Engineering \& BNRist, Tsinghua University, Beijing 100084, China\\
\textsuperscript{\rm 2}Microsoft Research Asia, Beijing 100080, China\\
  \{wuchuhan15, wufangzhao\}@gmail.com, \{xitwan, xing.xie\}@microsoft.com, yfhuang@tsinghua.edu.cn
}
\begin{document}

\maketitle

\begin{abstract}

News recommendation is important for online news services.
Existing news recommendation models are usually learned from  users' news click  behaviors.
Usually the behaviors of users with the same sensitive attributes (e.g., genders) have similar patterns and news recommendation models can easily capture these patterns. 
It may lead to some biases related to sensitive user attributes in the recommendation results, e.g., always recommending sports news to male users, which is unfair since users may not receive diverse news information.
In this paper, we propose a  fairness-aware news recommendation approach with  decomposed adversarial learning and orthogonality regularization, which can alleviate unfairness in news recommendation brought by the biases of sensitive user attributes.
In our approach, we propose to decompose the user interest model into two components.
One component aims to learn a bias-aware user embedding that captures the bias information on sensitive user attributes, and the other aims to learn a bias-free user embedding that only encodes attribute-independent user interest information for fairness-aware news recommendation.
In addition, we propose to apply an attribute prediction task to the bias-aware user embedding to enhance its ability on bias modeling, and we apply adversarial learning to the bias-free user embedding to remove the bias information from it.
Moreover, we propose an orthogonality regularization method to encourage the bias-free user embeddings to be orthogonal to the bias-aware one to better distinguish the bias-free user embedding from the bias-aware one.
For fairness-aware news ranking, we only use the bias-free user embedding.
Extensive experiments on benchmark dataset show that our approach can effectively improve fairness in news recommendation  with minor performance loss.
\end{abstract}

\section{Introduction}

\begin{figure}[!t]
    \centering
    \includegraphics[width=0.9\linewidth]{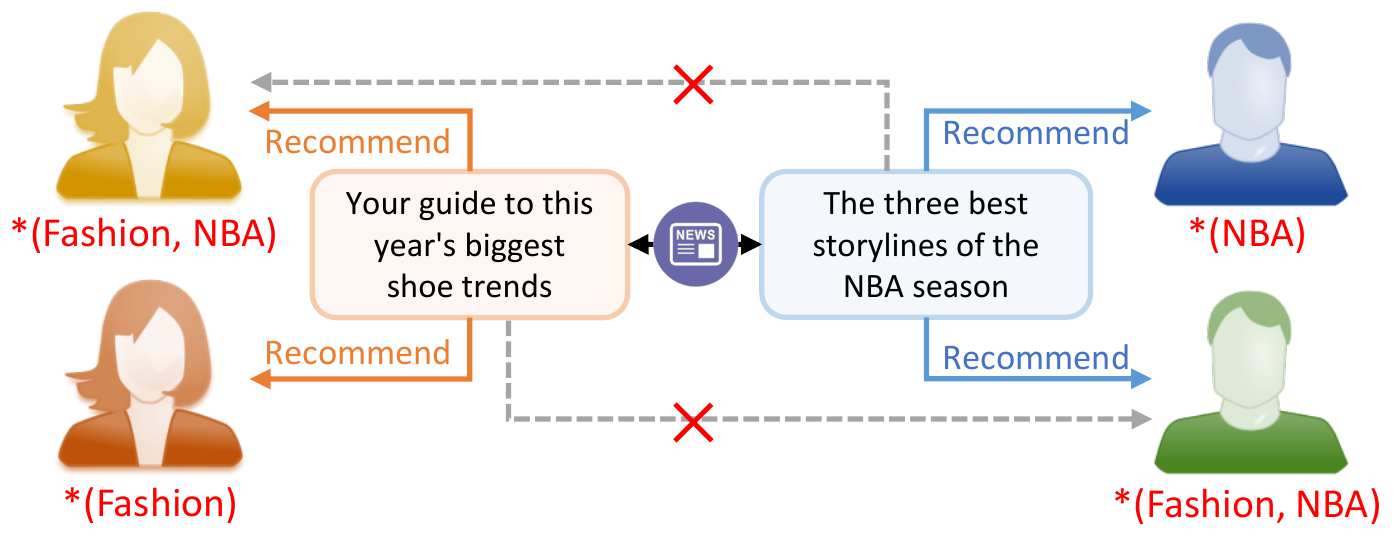}
    \caption{An example of gender bias in news recommendation. *Keywords under users represent their interest.}
    \label{fig:example}
\end{figure}

Personalized news recommendation techniques are critical for news websites to help users find their interested news and improve their reading experience~\cite{wu2019neural}.
Many existing methods for news recommendation rely on the news click behaviors of users to learn user interest models~\cite{okura2017embedding,wu2019npa}.
For example, \citeauthor{okura2017embedding}~\shortcite{okura2017embedding} proposed to learn user representations from the representations of clicked news articles with a GRU network.
\citeauthor{wu2019npa}~\shortcite{wu2019npa} proposed to use personalized attention networks to learn user representations from the representations of clicked news by using the embedding of user ID as attention query.
Usually, users with the same sensitive attributes (e.g., genders) may have similar patterns in their news click behaviors.
Taking user genders as an example,  in Fig.~\ref{fig:example} the female users may prefer fashion news while male users may prefer sports news.
However, user interest models can easily capture the these patterns and lead to some biases (e.g., gender bias) in the news recommendation results. 
For example, as shown in Fig.~\ref{fig:example}, since fashion news may be clicked by more female users while NBA news may be preferred more by male users, the model tends to only recommend fashion news to female users and NBA news to male users.
In this scenario, the recommendation results are heavily influenced by the biases brought by sensitive user attributes, and the users interested in both fashion and NBA cannot receive diverse news information, which is unfair and may be harmful for user experience.

In this paper, we propose a \underline{\textbf{fair}}ness-aware news \underline{\textbf{rec}}ommendation (FairRec) approach with decomposed adversarial learning and orthogonality regularization, which can effectively alleviate the unfairness in news recommendation brought by the biases related to sensitive user attributes like genders.
We propose  to decompose the user interest model into two components, where the first one aims to learn a bias-aware user embedding that captures biases related to sensitive user attributes from user behaviors, and the second one aims to learn a bias-free user embedding that mainly encodes attribute-independent user interest information for making fairness-aware news recommendation.
In addition, we apply a sensitive user attribute prediction task to the bias-aware user embedding to push it to convey more bias information, and we apply adversarial learning techniques to the bias-free user embedding to eliminate its information on sensitive user attributes.
Moreover, we propose an orthogonality regularization method to encourage the bias-free user embedding to be orthogonal to the bias-aware one, which can further remove the information  related to sensitive attributes from the bias-free user embedding.
To achieve fairness-aware news recommendation, we only use the bias-free user embedding  for personalized news ranking.
We conduct experiments on a benchmark news recommendation dataset, and the results show that our approach can effectively improve news recommendation fairness with acceptable performance sacrifice.

The major contributions of this paper include:
\begin{itemize}
    \item This is the first work that explores to improve fairness in news recommendation by proposing a fairness-aware news recommendation framework.
    \item We propose a decomposed adversarial learning method with orthogonality regularization to learn bias-free user embeddings for fairness-aware news ranking.
    \item Extensive experiments on real-world dataset demonstrate that our approach can effectively improve fairness in news recommendation.
\end{itemize}
\section{Related Work}

\subsection{News Recommendation}
News recommendation is an essential technique for online news platform to provide personalized news services.
Accurately modeling of user interest is critical for news recommendation~\cite{wu2019npa}.
In many existing news recommendation methods,  the interest of users is modeled by their news click behaviors~\cite{wang2018dkn,wu2019neuralnaml,zhu2019dan,an2019neural,wu2019npa,wu2019neuralnrms,qi2020privacy,wang2020fine,hu2020graph}.
For example, \citeauthor{okura2017embedding}~\shortcite{okura2017embedding} proposed to use a GRU network to learn user representations from the representations of clicked news.
\citeauthor{wang2018dkn}~\shortcite{wang2018dkn} proposed to learn user representations based on the relevance between the representations of clicked and candidate news. 
\citeauthor{wu2019neuralnrms}~\shortcite{wu2019neuralnrms} proposed to learn user representations from clicked news via multi-head self-attention networks.
These existing methods usually learn news recommendation models from users' news click behaviors.
However, their models can easily grasp the similar patterns in the  behaviors of users with the same sensitive attributes and lead to biased news recommendation results.
Thus, the users may not receive diverse news information, which is harmful to user experience.
Different from these methods, in our approach we propose a decomposed adversarial learning approach with orthogonality regularization to learn bias-free user embeddings for fairness-aware news ranking, which can substantially improve news recommendation fairness with small performance sacrifice.

\subsection{Fairness-aware Recommendation}

The problem of fairness in recommendation has attracted much attention in recent years~\cite{beutel2019fairness,ekstrand2019fairness,fu2020fairness,patro2020fairrec}.
Some studies explore the problem of provider-side fairness, e.g., items from different providers have a fair chance of being recommended~\cite{lee2014fairness,kamishima2014correcting,liu2019personalized}.
There are also several methods that address the problem of customer-side fairness, e.g., provide similar recommendations for users with different sensitive attributes~\cite{xiao2017fairness,zhu2018fairness,burke2018balanced}.
Many methods study customer-side fairness on e-commerce scenarios by using ratings to indicate fairness~\cite{yao2017beyond}.
For example, \citeauthor{yao2017beyond}~\shortcite{yao2017beyond} proposed four different metrics based on the predicted and real ratings of users with different attributes to measure unfairness.
They proposed to regularize collaborative filtering models with one of the unfairness metrics to explore the model performance in minimizing each form of unfairness.
\citeauthor{farnadi2018fairness}~\shortcite{farnadi2018fairness} proposed to use probabilistic soft logic (PSL) rules to balance the ratings for both users in different groups by un-biasing the ratings for each item.
These methods mainly aim to balance the recommendation performance for users with different sensitive attributes. 
\citeauthor{geyik2019fairness}~\shortcite{geyik2019fairness}  explored several re-ranking rules to provide fair rankings of LinkedIn users based on their ranking scores and the desired proportions over different user attributes.
This method aims to provide fair rankings of users with different attributes.
Different from these methods, our approach focuses on the fairness of news recommendation results rather than accuracy, and we need to rank news rather than users.
We propose a decomposed adversarial learning method to learn bias-free user embeddings, which is used to generate fairness-aware news recommendation results.

\section{Methodology}

In this section, we first present the problem definitions of this paper, then introduce  the details of our fairness-aware news recommendation framework  with decomposed adversarial learning and orthogonality regularization.

\subsection{Problem Definition}

For a target user $u$ with the sensitive attribute $z$, we assume that she has clicked  $N$ news articles, which are denoted as $\mathcal{D}=\{D_1, D_2, ..., D_N\}$.
We denote the candidate news set for this user as $\mathcal{D}^c=\{D^c_1, D^c_2, ..., D^c_M\}$, where $M$ is the number of candidate news.
The gold click labels of the target user $u$ clicking these candidate news are denoted as $[y_1, y_2, ..., y_M]$.
The click labels predicted by the  news recommendation model are denoted as $[\hat{y}_1, \hat{y}_2, ..., \hat{y}_M]$.
Candidate news are sorted by these predicted click labels, and the top $K$ ranked candidate news set (regarded as the recommendation result) is denoted as $\mathcal{D}^r=\{D^c_{i_1}, D^c_{i_2}, ..., D^c_{i_K}\}$.
The unfairness of the recommendation result $\mathcal{D}^r$ is defined as how discriminative it is for inferring the sensitive user attribute $z$.
If $z$ can be predicted from $\mathcal{D}^r$ more accurately, the recommendation result is more unfair since it is  more heavily influenced by the sensitive user attribute.

\begin{figure*}[!t]
    \centering
    \includegraphics[width=0.68\linewidth]{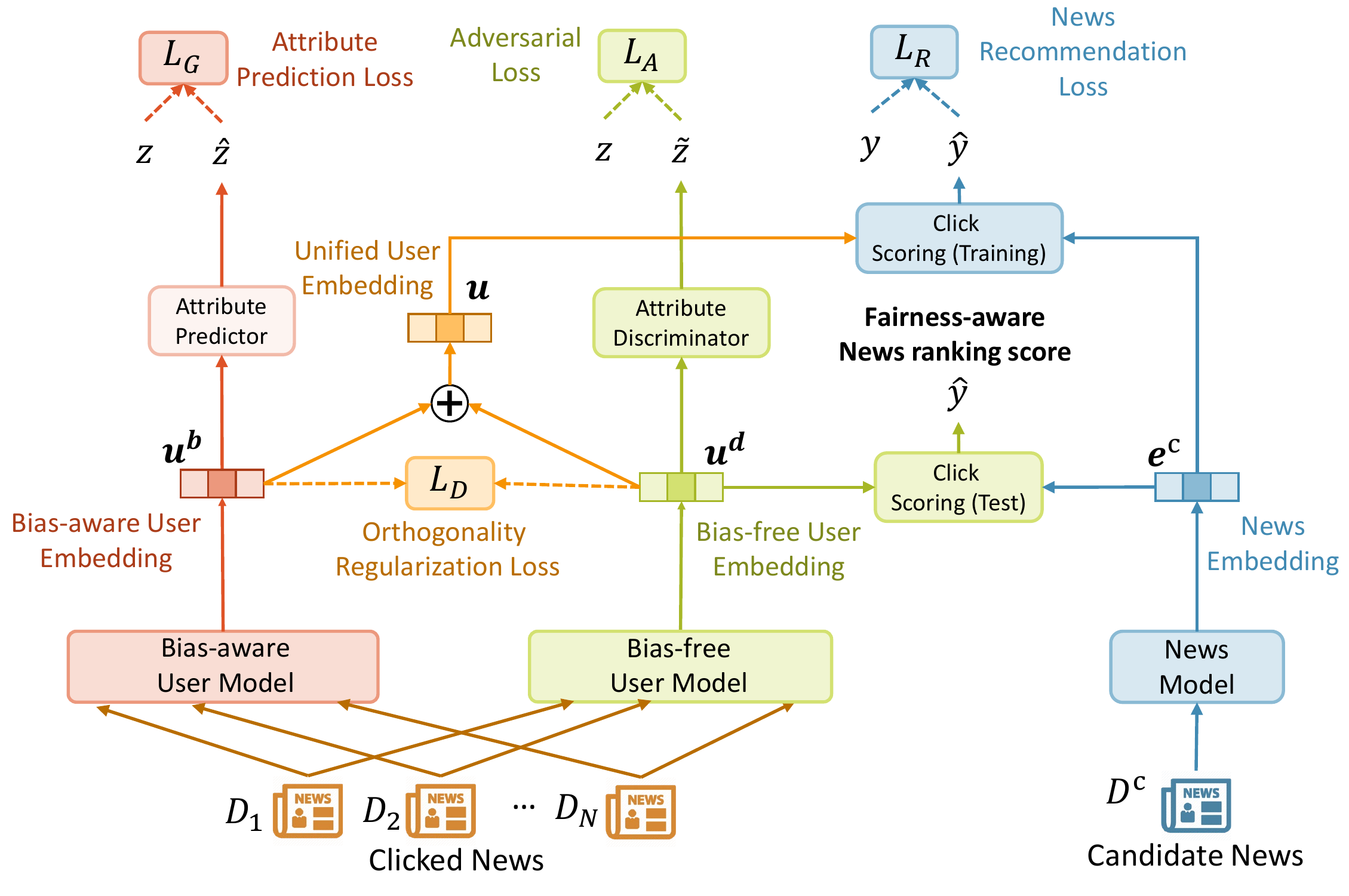}
    \caption{The architecture of our \textit{FairRec} approach.}
    \label{fig:model}
\end{figure*}

\subsection{Framework of FairRec}

First, we introduce the framework of the proposed \underline{\textbf{fair}}ness-aware news \underline{\textbf{rec}}ommendation (FairRec) method, as shown in Fig.~\ref{fig:model}.
It  mainly aims to compute a fairness-aware news ranking score for each candidate news of a user, which is further used to rank candidate news and generate fairness-aware news recommendation results for this user.
More specifically, our \textit{FairRec} framework uses a news model to learn the embeddings of candidate news, a bias-free user model to learn the bias-free embeddings of users which minimally contain the bias information on the sensitive user attribute, and a click scoring model to compute the fairness-aware news ranking scores based on the  bias-free user embedding and candidate news embeddings.
We briefly introduce these components as follows.

The news and user models in our approach are based on those in the NRMS~\cite{wu2019neuralnrms} method.
The news model learns news representations from news titles.
It first uses a multi-head self-attention network to capture the contexts of words within a news title, and then uses an attentive pooling network to learn news representations by modeling the importance of different words.
We denote the representation of the candidate news $D^c$ learned by the news model as $\mathbf{e}^c$.
The user model learns the representation of a target user $u$ from her clicked news $[D_1, D_2, ..., D_N]$.
It first uses a news model to learn the representations of these clicked news, then uses a combination of multi-head self-attention network and attentive pooling network to obtain the unified user representations.
We denote the bias-free user embedding learned by this user model as $\mathbf{u}^d$.
Finally, the click scoring module computes the fairness-aware ranking score $\hat{y}$ based on the bias-free user embedding $\mathbf{u}^d$ and the candidate news embedding $\mathbf{e}^c$.
Following many previous methods~\cite{okura2017embedding,wu2019npa}, we use the dot product function to compute the fairness-aware ranking score by evaluating the relevance between the bias-free user embedding and candidate news embedding, i.e., $\hat{y}=\mathbf{u}^d\cdot \mathbf{e}^c$.
The ranking scores of candidate news are further used for personalized news ranking and display.

\subsection{Decomposed Adversarial Learning with Orthogonality Regularization}

Then, we introduce the details of the proposed decomposed adversarial learning and orthogonality regularization method for learning bias-free user embeddings.
In our fairness-aware recommendation framework, a core problem is how to learn the bias-free user embedding $\mathbf{u}^d$ from users' news click behaviors.
However, since the users with the same sensitive attribute usually have some similar patterns in their news click behaviors, the user model can easily capture these patterns from users' news click behaviors and generate biased user embeddings.
Thus, it is non-trivial to learn bias-free user embeddings from the biased user behaviors. 

Adversarial learning is a technique that can be used to learn bias-free deep representations from biased data~\cite{madras2018learning,elazar2018adversarial}.
Its mission is to enforce the deep representations to be maximally informative for predicting the labels of the main task, and meanwhile to be minimally discriminative for predicting sensitive attributes~\cite{du2019fairness}.
Thus, adversarial learning has the potential to learn bias-free user embeddings by removing the bias information about sensitive user attributes.
A straightforward way is to apply an attribute discriminator to  the user embeddings learned by the user model to infer the sensitive user attribute, and penalize the model according to the negative gradients from the adversarial loss that indicates the informativeness of user embeddings for sensitive user attribute prediction.
At the same time, the user embeddings are also used to evaluate the relevance between the user and candidate news for news recommendation model training.
Unfortunately, users' sensitive attributes may be informative for the main news recommendation task, and the bias information related to the sensitive user attribute may be encoded into the user embeddings, making it difficult to be removed by  adversarial learning.
As an alternate, we propose to decompose the user interest model into two components, i.e., a bias-aware one that mainly aims to learn bias-aware user embeddings that capture the bias information on sensitive user attributes, and a bias-free one that only encodes the attribute-independent information of user interest into bias-free user embeddings.
To push the bias-aware user embedding to be more attribute-discriminative, we propose to apply a sensitive attribute prediction task to the bias-aware user embedding.
The user attribute $z$ is predicted by an attribute predictor as follows\footnote{We assume the attribute is a categorical variable here.}:
\begin{equation}
    \hat{z}=\mathrm{softmax}(\mathbf{W}^b\mathbf{u}^b+\mathbf{b}^b),
\end{equation}
where $\mathbf{W}^b$ and $\mathbf{b}^b$ are parameters, $\hat{z}$ is the predicted probability vector.
The loss function for attribute prediction is crossentropy, which is formulated as:
\begin{equation}
    \mathcal{L}_G=-\frac{1}{U}\sum_{j=1}^U\sum_{i=1}^C z^j_i\log(\hat{z}^j_i),
\end{equation}
where $z^j_i$ and $\hat{z}^j_i$ respectively stand for the gold and predicted probability of the $j$-th user's attribute in the $i$-th class, and $U$ is the number of users.

Usually, the supervision of the main recommendation task may also encode the bias information about sensitive user attribute into the bias-free user embedding.
Thus, in order to eliminate the bias information, we propose to apply adversarial learning to the bias-free user embedding.
More specifically, we use a attribute discriminator  to predict user attributes according to the bias-free user embedding as follows:
\begin{equation}
    \tilde{z}=\mathrm{softmax}(\mathbf{W}^d\mathbf{u}^d+\mathbf{b}^d),
\end{equation}
where $\mathbf{W}^d$ and $\mathbf{b}^d$ are parameters. 
The adversarial loss function of the discriminator is similar to the attribute predictor, which is formulated as follows:
\begin{equation}
    \mathcal{L}_A=-\frac{1}{U}\sum_{j=1}^U\sum_{i=1}^C z^j_i\log(\tilde{z}^j_i).
\end{equation}
To avoid the discriminator from inferring user attributes from the bias-free user embedding, we use the negative gradients of the discriminator to penalize the model.

Unfortunately, the bias-free user embedding may still contain some information related to the sensitive user attribute.
This is because the discriminator usually cannot perfectly infer the sensitive user attribute, and there are shifts between the decision boundary of the discriminator and the real distribution of the sensitive user attribute.
Since the bias-free user embedding generated by the user model only needs to cheat the discriminator, it does not necessarily fully remove the information of sensitive user attributes.
To solve this problem, we propose an orthogonality regularization method to further purify the bias-free user embedding.
Concretely, it regularizes the bias-aware user embedding and bias-free user embedding by encouraging them to be orthogonal to each other.
The regularization loss function is formulated as follows:
\begin{equation}
    \mathcal{L}_D=\frac{1}{U}\sum_{i=1}^U|\frac{\mathbf{u}^b_i\cdot\mathbf{u}^d_i}{||\mathbf{u}^b_i||\cdot||\mathbf{u}^d_i||}|,
\end{equation}
where $\mathbf{u}^b_i$ and $\mathbf{u}^d_i$ are respectively the bias-aware and bias-free embeddings of the $i$-th user.

\subsection{Model Training}

Finally, we introduce how to train the models in our approach.
In our \textit{FairRec} framework, the bias-aware user embedding mainly contains the information on sensitive user attribute, and the bias-free user embedding mainly encodes attribute-independent user interest information. 
The information in both embeddings is correlated with the main recommendation task.\footnote{In fact, bias-independent user interest information may also exist in both kinds of user embeddings. We will explore how to push the bias-independent user interests to be maximally captured by the bias-free user embedding in our future work.}
Thus, we add both user embeddings together to form a unified one for training the recommendation model, i.e., $\mathbf{u}=\mathbf{u}^b+\mathbf{u}^d$.
We denote the representation of the candidate news $D^c$ as $\mathbf{e}^c$, which is encoded by the news model.
The probability of a user $u$ clicking news $D^c$ is predicted by $\hat{y}=\mathbf{u}\cdot \mathbf{e}^c$.
Following~\cite{huang2013learning,wu2019neuralnrms}, we use negative sampling techniques to construct labeled samples for news recommendation model training.
For each candidate news clicked by a user, we randomly sample $T$ negative news in the same session which are not clicked.
The loss function for news recommendation is the negative log-likelihood of the posterior click probability of clicked news, which is formulated as follows:
\begin{equation}
    \mathcal{L}_R=-\frac{1}{N_c}\sum_{i=1}^{N_c}\log[\frac{\exp(\hat{y}_i)}{\exp(\hat{y}_i)+\sum_{j=1}^T\exp(\hat{y}_{i,j})}],
\end{equation}
where $\hat{y}_i$ and $\hat{y}_{i,j}$ are the click scores of the $i$-th clicked candidate news and its associated $j$-th negative news, respectively.
$N_c$ is the number of clicked candidate news for training.
The entire framework is trained collaboratively, and the final loss function for the recommendation model (except the discriminator) is a weighted summation of the news recommendation, attribute prediction, orthogonality regularization and adversarial loss functions, which is formulated as follows:
\begin{equation}
    \mathcal{L}=\mathcal{L}_R+\lambda_G\mathcal{L}_G+\lambda_D\mathcal{L}_D-\lambda_A\mathcal{L}_A, \label{eq1}
\end{equation}
where $\lambda_G$, $\lambda_D$ and $\lambda_A$ are coefficients that control the importance of their corresponding losses.

\section{Experiments}
\subsection{Dataset and Experimental Settings}

In our experiments, we focus on gender parity in validating the effectiveness of our fairness-aware news recommendation approach.
The dataset used in our experiments is provided by~\cite{wu2019neural}, which contains the news impression logs of users and their gender labels (if available).
It contains 10,000 users and their news browsing behaviors (from Dec. 13, 2018 to Jan. 12, 2019), and 4,228 users provide their gender label (2,484 male users and 1,744 female users).
For the users without gender labels, the attribute prediction and adversarial losses are deactivated.
The logs in the last week are used for test, and the rest are used for model training.
In addition, we randomly sample 10\% of training logs for validation.
The statistics of this dataset are summarized in Table~\ref{dataset}.

\begin{table}[t]
\centering
	\resizebox{1.0\linewidth}{!}{
\begin{tabular}{lrlr}
\Xhline{1.5pt}
\#users       & 10,000  & avg. \#words per news title & 11.29     \\
\#news        & 42,255  & \#clicked news logs         & 503,698   \\
\#impressions & 360,428 & \#non-clicked news logs     & 9,970,795 \\ \Xhline{1.5pt}
\end{tabular}
}
\caption{Statistics of the dataset.}\label{dataset}
\end{table}

\begin{table*}[t]
	\centering
\resizebox{0.9\textwidth}{!}{
\begin{tabular}{ccccccccc}
\Xhline{1.5pt} 
\multirow{2}{*}{\textbf{Methods}} & \multicolumn{2}{c}{Top 1}                                                           & \multicolumn{2}{c}{Top 3}                                                           & \multicolumn{2}{c}{Top 5}                                                           & \multicolumn{2}{c}{Top 10}                                                          \\ \cline{2-9} 
                                  & Accuracy                                 & Macro-F                                  & Accuracy                                 & Macro-F                                  & Accuracy                                 & Macro-F                                  & Accuracy                                 & Macro-F                                  \\ \hline
LibFM                             & 59.78$\pm$0.64                           & 59.34$\pm$0.62                           & 63.25$\pm$0.61                           & 63.04$\pm$0.60                           & 64.63$\pm$0.59                           & 64.46$\pm$0.56                           & 66.42$\pm$0.54                           & 66.25$\pm$0.51                           \\
EBNR                              & 61.65$\pm$0.70                           & 61.31$\pm$0.67                           & 65.40$\pm$0.64                           & 65.12$\pm$0.64                           & 66.86$\pm$0.61                           & 66.72$\pm$0.60                           & 68.65$\pm$0.51                           & 68.49$\pm$0.50                           \\
DKN                               & 61.88$\pm$0.74                           & 61.54$\pm$0.71                           & 65.84$\pm$0.67                           & 65.61$\pm$0.66                           & 67.33$\pm$0.63                           & 67.19$\pm$0.63                           & 69.12$\pm$0.56                           & 68.98$\pm$0.55                           \\
DAN                               & 62.54$\pm$0.72                           & 62.29$\pm$0.70                           & 66.22$\pm$0.70                           & 65.97$\pm$0.69                           & 67.96$\pm$0.67                           & 67.79$\pm$0.66                           & 69.74$\pm$0.54                           & 69.57$\pm$0.52                           \\
NPA                               & 62.67$\pm$0.68                           & 62.31$\pm$0.67                           & 66.43$\pm$0.67                           & 66.13$\pm$0.65                           & 68.07$\pm$0.64                           & 67.84$\pm$0.62                           & 69.85$\pm$0.52                           & 69.62$\pm$0.49                           \\
NRMS                              & 63.13$\pm$0.71                           & 62.75$\pm$0.70                           & 66.89$\pm$0.68                           & 66.54$\pm$0.66                           & 68.32$\pm$0.67                           & 67.96$\pm$0.65                           & 70.12$\pm$0.59                           & 69.94$\pm$0.56                           \\ \hline
MR                                & 60.75$\pm$0.76                           & 60.55$\pm$0.73                           & 63.27$\pm$0.67                           & 62.98$\pm$0.64                           & 65.45$\pm$0.68                           & 65.23$\pm$0.65                           & 67.24$\pm$0.60                           & 67.01$\pm$0.57                           \\
AL                                & 58.86$\pm$0.75                           & 58.51$\pm$0.73                           & 62.67$\pm$0.65                           & 62.41$\pm$0.63                           & 64.92$\pm$0.63                           & 64.61$\pm$0.61                           & 66.70$\pm$0.54                           & 66.39$\pm$0.52                           \\
ALGP                              & 57.93$\pm$0.71                           & 57.64$\pm$0.70                           & 61.84$\pm$0.66                           & 61.62$\pm$0.65                           & 63.73$\pm$0.61                           & 63.52$\pm$0.60                           & 65.52$\pm$0.51                           & 65.30$\pm$0.49                           \\ \hline
FairRec                               & \textbf{51.11}$\pm$0.69 & \textbf{50.99}$\pm$0.66 & \textbf{52.20}$\pm$0.61 & \textbf{52.06}$\pm$0.60 & \textbf{52.83}$\pm$0.54 & \textbf{52.61}$\pm$0.54 & \textbf{53.40}$\pm$0.48 & \textbf{53.12}$\pm$0.46 \\ \hline
Random                            & 50.11$\pm$0.30                           & 50.09$\pm$0.28                           & 50.04$\pm$0.21                           & 50.03$\pm$0.20                           & 50.06$\pm$0.17                           & 50.03$\pm$0.16                           & 50.02$\pm$0.14                           & 50.01$\pm$0.10                           \\  
\Xhline{1.5pt}
\end{tabular}
}
	\caption{News recommendation fairness of different methods. Lower scores indicate better fairness. The best results except random ranking are in bold.}\label{table.result}

\end{table*}

\begin{table}[t]
	\centering
	\resizebox{0.45\textwidth}{!}{
\begin{tabular}{ccccc}
\Xhline{1.5pt}
\textbf{Methods} &               AUC            & MRR            & nDCG@5         & nDCG@10        \\ \hline
LibFM                & 56.83$\pm$0.51          & 24.20$\pm$0.53         & 26.95$\pm$0.49                           & 35.64$\pm$0.52                            \\
EBNR                 & 60.94$\pm$0.24          & 28.22$\pm$0.25          & 30.31$\pm$0.23                           & 39.60$\pm$0.24                            \\ 
DKN                & 60.34$\pm$0.33          & 27.51$\pm$0.29          & 29.75$\pm$0.31                           & 38.79$\pm$0.30                            \\
DAN               & 61.43$\pm$0.31          & 28.62$\pm$0.30          & 30.66$\pm$0.32                           & 39.81$\pm$0.33                            \\
NPA              & 62.33$\pm$0.25          & 29.46$\pm$0.23          & 31.57$\pm$0.22                           & 40.71$\pm$0.23                            \\
NRMS                & 62.89$\pm$0.22          & 29.93$\pm$0.20          & 32.19$\pm$0.18                           & 41.28$\pm$0.18                            \\ \hline                                  
FairRec            &  61.95$\pm$0.22          & 29.01$\pm$0.21         & 31.25$\pm$0.18                           & 40.24$\pm$0.21                           \\
\Xhline{1.5pt}
\end{tabular}
}
\caption{News recommendation performance of different methods. Higher scores indicate better results.}\label{table.result2}
\end{table}

In our experiments, pre-trained Glove~\cite{pennington2014glove} embeddings are used to initialize the word embeddings. 
Adam~\cite{kingma2014adam} is used as the model optimizer, and the learning rate is 0.001.
The dropout~\cite{srivastava2014dropout} ratio is 0.2.
The loss coefficients in Eq.~(\ref{eq1}) are all set to 0.5.
These hyperparameters are tuned on the validation set.

Since the problem studied in this paper is the fairness of recommendation results rather than accuracy~\cite{he2020geometric}, the fairness metrics based on user ratings used in several existing methods~\cite{yao2017beyond,farnadi2018fairness} may not be suitable.
To quantitatively measure the fairness of news recommendation results, we propose to use the prediction performance of sensitive user attribute based on the top $K$ ranked candidate news in each session as the indication of recommendation fairness.
The attribute prediction model contains a user model to learn user embeddings and an attribute predictor with a dense layer to infer the attributes. 
Since the dataset has an imbalanced gender distribution and there are system gender biases in the impression logs brought by news recall and pre-ranking, we build a new dataset from the original dataset to better evaluate recommendation fairness.
We down-sample the number of male users to balance user gender, and use the entire news set as the candidate news set $\mathcal{D}^c$ for ranking to avoid impression gender bias.
We use 80\% of users for training the attribute prediction model, 10\% for validation and the rest 10\% for test.
Following~\cite{wu2019neural}, we use accuracy and macro F-score as the metrics to indicate fairness, where lower scores mean better recommendation fairness.
To evaluate the performance of news recommendation, we use the average AUC, MRR, nDCG\@5 and nDCG\@10 scores of test sessions.
We independently repeat each experiment 10 times and report the average results with standard deviations.

\subsection{Performance Evaluation}
In this section, we evaluate the performance of our \textit{FairRec} approach in terms of fairness and news recommendation.
We compare \textit{FairRec} with various baseline methods for news recommendation, including:
(1) LibFM~\cite{rendle2012factorization}, a popular recommendation tool based on factorization machine;
(2) EBNR~\cite{okura2017embedding}, an embedding-based news recommendation method that employs autoencoders to learn news representations and a GRU network to generate user representations;
(3) DKN~\cite{wang2018dkn}, using knowledge-aware CNNs to encode news representations and the relevance between representations of clicked news and candidate news to build user representations;
(4) DAN~\cite{zhu2019dan}, using CNN to learn news representations and attentive LSTM to form user representations;
(5) NPA~\cite{wu2019npa}, using personalized attention networks to learn news and user representations;
(6) NRMS~\cite{wu2019neuralnrms}, using a combination of multi-head self-attention and additive attention to learn news and user representations.
In addition, we compare the recommendation fairness of several additional methods, including:
(7) MR~\cite{yao2017beyond}, using an unfairness loss to regularize our recommendation model. 
We regard the predicted click scores as ``ratings'';
(8) AL~\cite{wadsworth2018achieving}, applying adversarial learning to the single user embedding;
(9) ALGP~\cite{zhang2018mitigating}, using gradients projection in adversarial learning.
(10) Random, ranking candidate news randomly, which is used to show the ideal recommendation fairness.
The recommendation fairness of different methods under $K=1, 3, 5$ or 10 and their recommendation performance are respectively shown in Tables~\ref{table.result} and~\ref{table.result2}.
From the results, we have several observations.

First, compared with random ranking, the recommendation results of most methods are biased.
This is possibly because users with the same attributes such as demographics usually have similar patterns in their behaviors, and user models may inherit  these biases and encode them into the news ranking results.
Second, compared with the methods that do not consider the fairness of recommendation (e.g., DAN, NPA and NRMS), fairness-aware methods  (MR, AL, ALGP and FairRec) yield better recommendation fairness.
Among them, the methods based on adversarial learning techniques perform better than the model regularization (MR) method that uses an unfairness loss to regularize the model.
It shows that adversarial learning is more effective in improving the fairness of recommendation results by reducing the bias information in user embeddings.
Third, compared with AL and ALGP, our approach achieves better recommendation fairness with a substantial margin.
This may be because in AL and ALGP there are shifts between the decision boundaries of their discriminators and the real attribute distributions.
Since the bias-free user embeddings only need to deceive the discriminator, they may not be orthogonal to the space of sensitive user attribute, which means that the bias information is not fully removed.
Our approach uses a decomposed adversarial learning method with orthogonality regularization, which can learn bias-free user embeddings more effectively.
Fourth, our approach can effectively improve recommendation fairness and meanwhile keep good recommendation performance.
Compared with random ranking, our approach can almost achieve comparable recommendation fairness under different $K$.
In addition,  the recommendation performance of our approach is quite competitive.
It outperforms several strong baseline methods like DKN and DAN, and the performance sacrifice is not large compared with its basic model NRMS that does not consider recommendation fairness.
These results validate that our approach can effectively improve fairness in news recommendation with minor performance loss.

\begin{figure}[!t]
    \centering
    \includegraphics[width=0.75\linewidth]{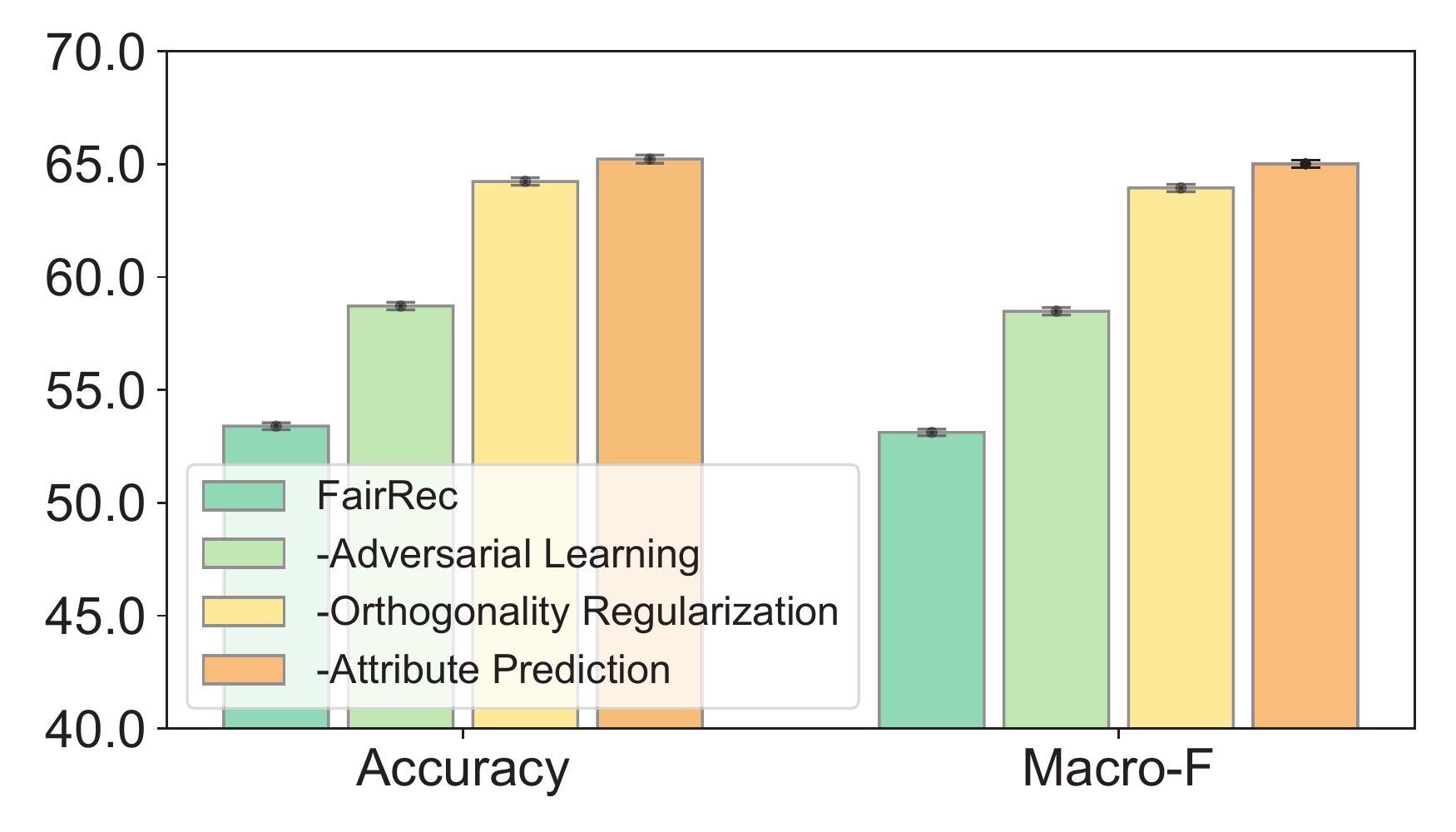}
    \caption{The effectiveness of decomposed adversarial learning. Lower scores represent better fairness.}
    \label{fig:effect}
\end{figure}

\subsection{Effectiveness of Decomposed Adversarial Learning}

In this section, we conduct several ablation studies to verify the effectiveness of the core components in our \textit{FairRec} approach, i.e., attribute prediction, adversarial learning and orthogonality regularization.
We compare the recommendation fairness  (under $K=10$) of \textit{FairRec} and its variants with one of these components removed, and the results are illustrated in Fig.~\ref{fig:effect}.
We have several findings from this plot.
First, applying the attribute prediction task to the bias-aware user embedding is very important.
This is because the attribute prediction task can greatly enhance the ability of bias-aware user embedding in bias modeling, which can help further remove the bias information from the bias-free user embedding. 
Second, applying adversarial learning to the bias-free user embedding is helpful for improving the fairness of news recommendation.
This is because adversarial learning can encourage the bias-free user embedding to minimize the information for inferring the sensitive user attributes.
Third, the orthogonality regularization added to the bias-aware and bias-free user embeddings can also effectively improve the recommendation fairness.
It is because that this auxiliary regularization can push the bias-free user embedding to be orthogonal to the bias-aware user embedding and hence contains less bias information on sensitive user attributes. 

\subsection{Hyperparameter Analysis}

\begin{figure}[!t]
    \centering
    \subfigure[Fairness.]{
    \includegraphics[width=0.46\linewidth]{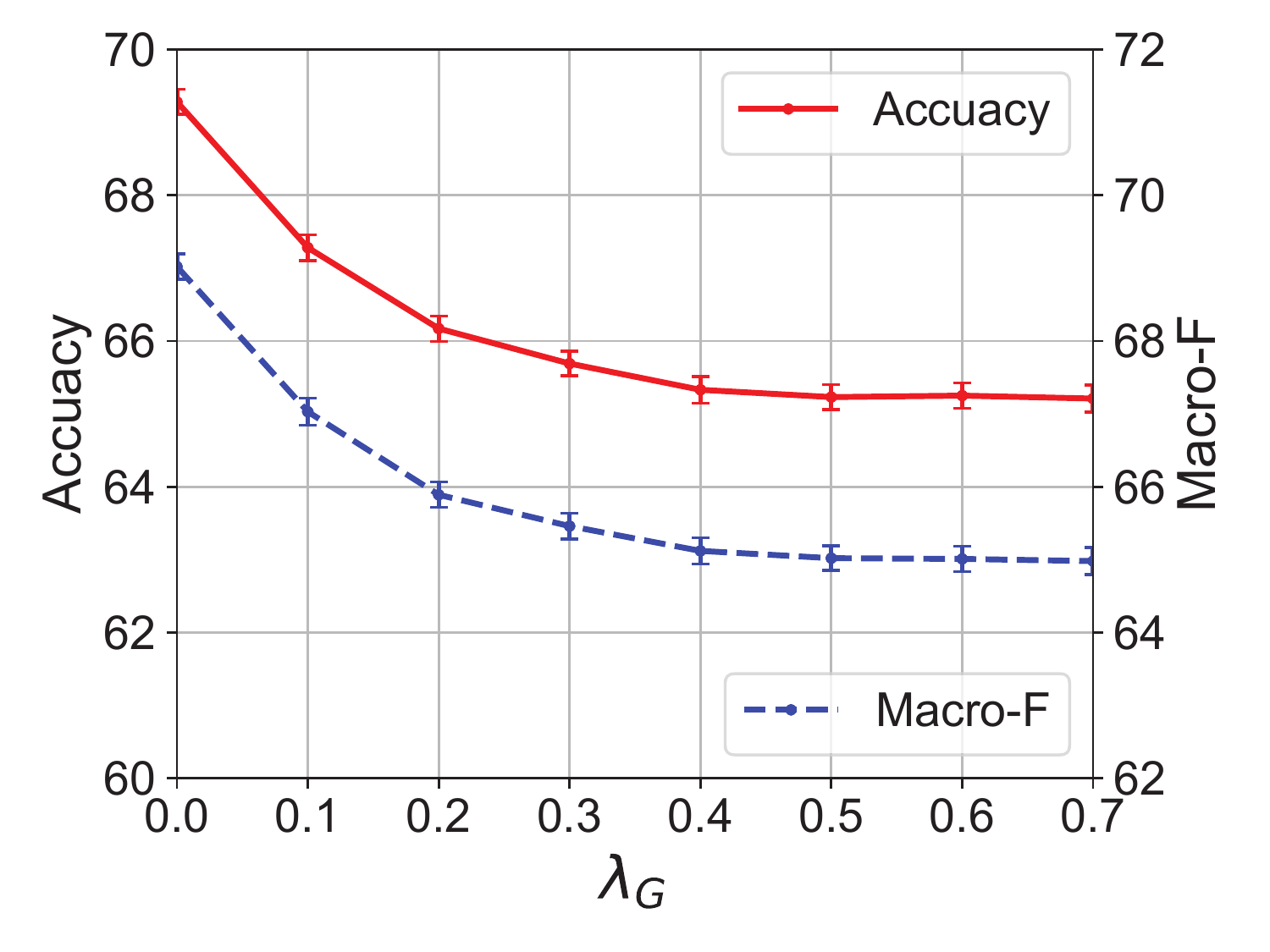}\label{g1}
    }
        \subfigure[Performance.]{
    \includegraphics[width=0.46\linewidth]{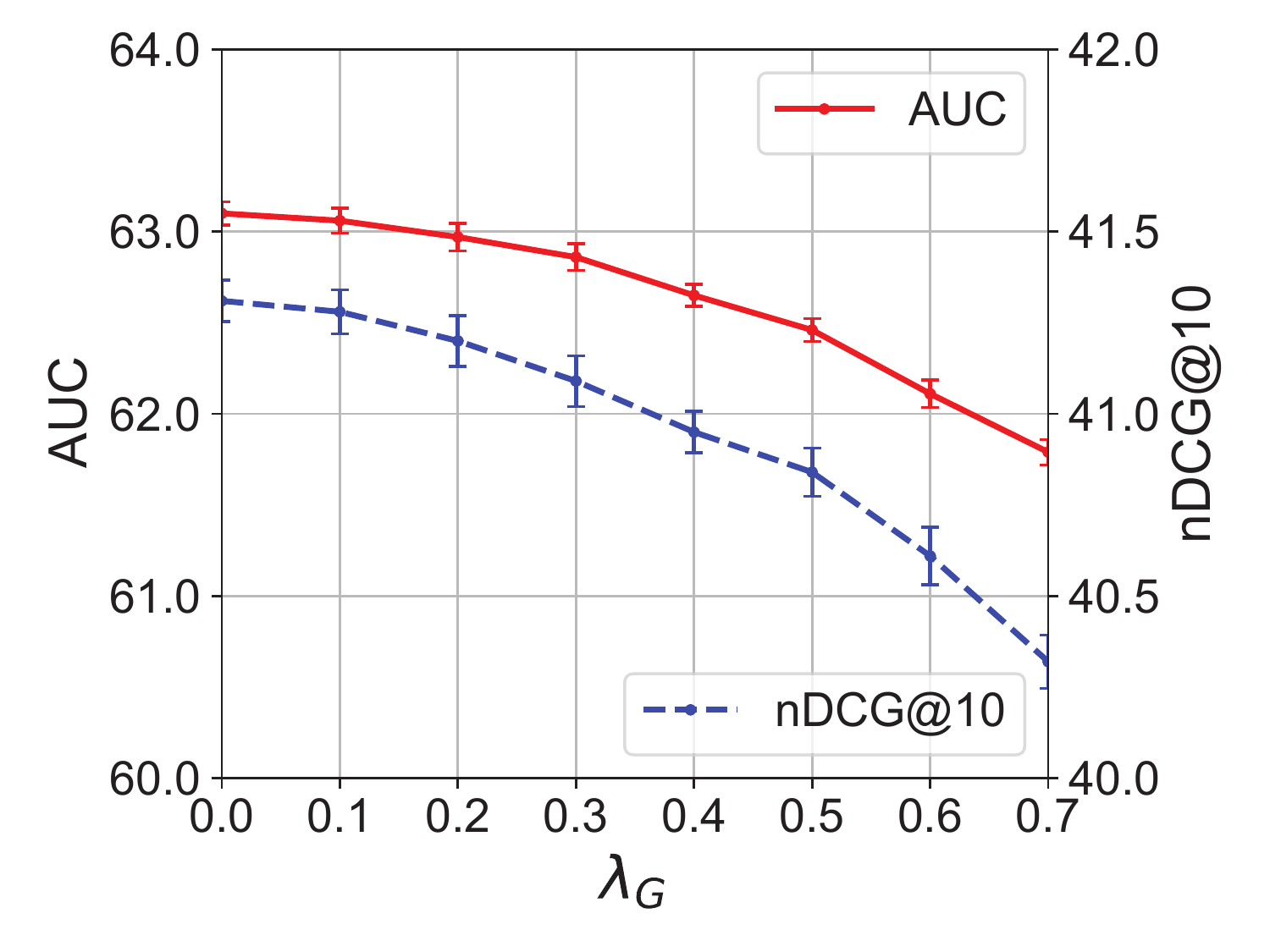}\label{g2}
    }
    \caption{The news recommendation fairness and performance w.r.t. different $\lambda_G$.}
    \label{fig:lambdag}
\end{figure}

\begin{figure}[!t]
    \centering
    \subfigure[Fairness.]{
    \includegraphics[width=0.46\linewidth]{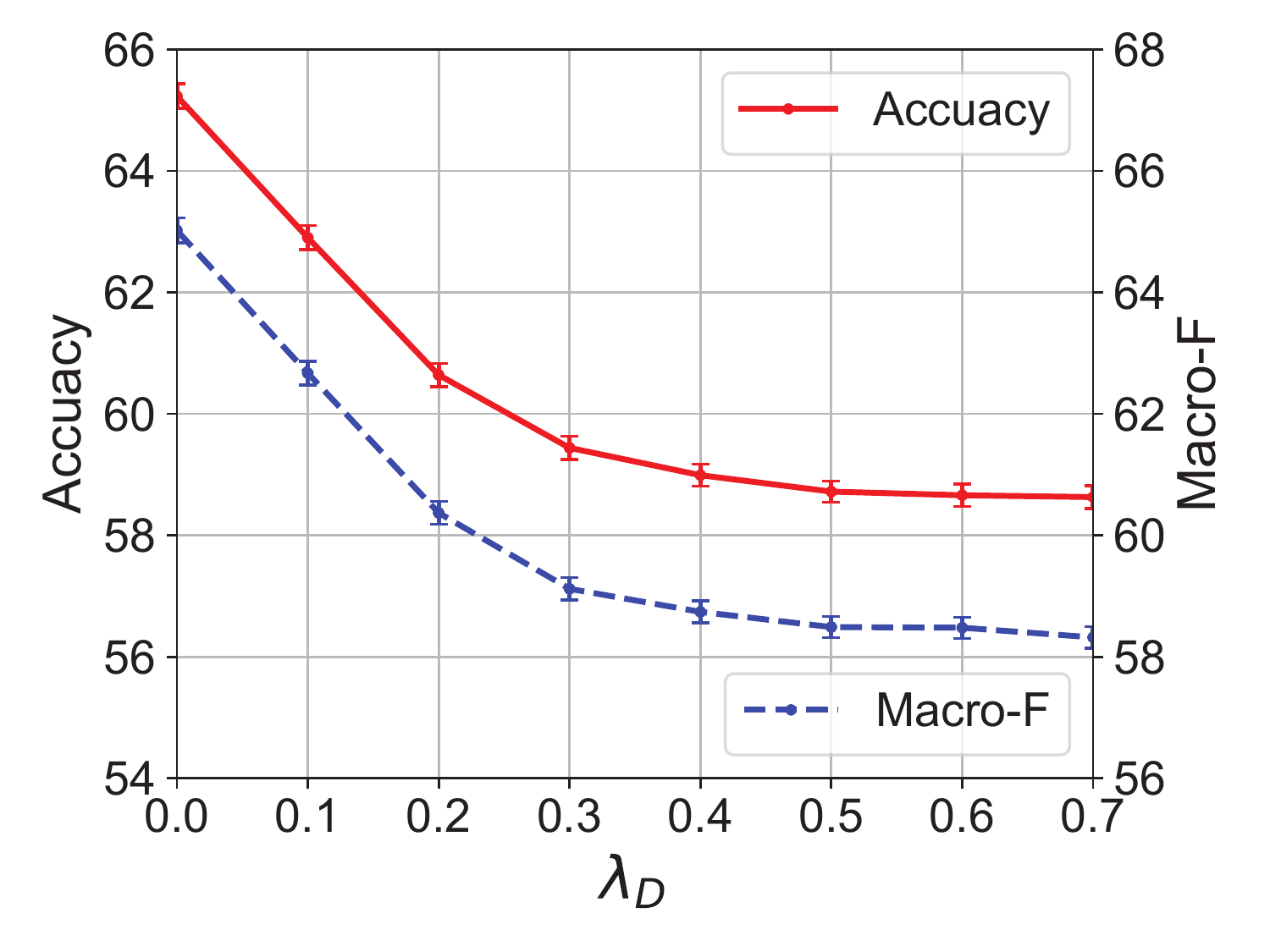}\label{d1}
    }
        \subfigure[Performance.]{
    \includegraphics[width=0.46\linewidth]{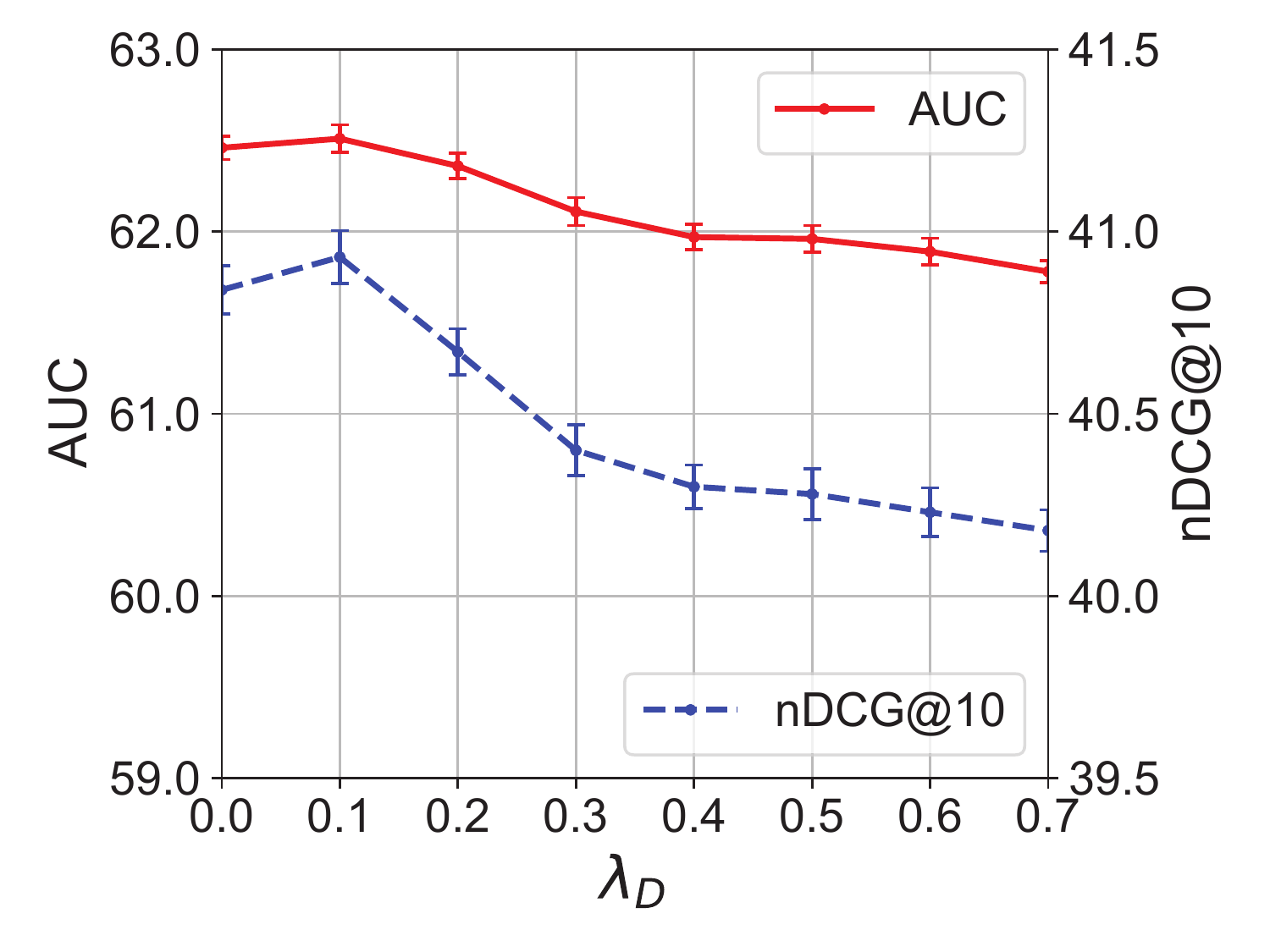}\label{d2}
    }
    \caption{The news recommendation fairness and performance w.r.t. different $\lambda_D$.}
    \label{fig:lambdag}
\end{figure}

\begin{figure}[!t]
    \centering
    \subfigure[Fairness.]{
    \includegraphics[width=0.46\linewidth]{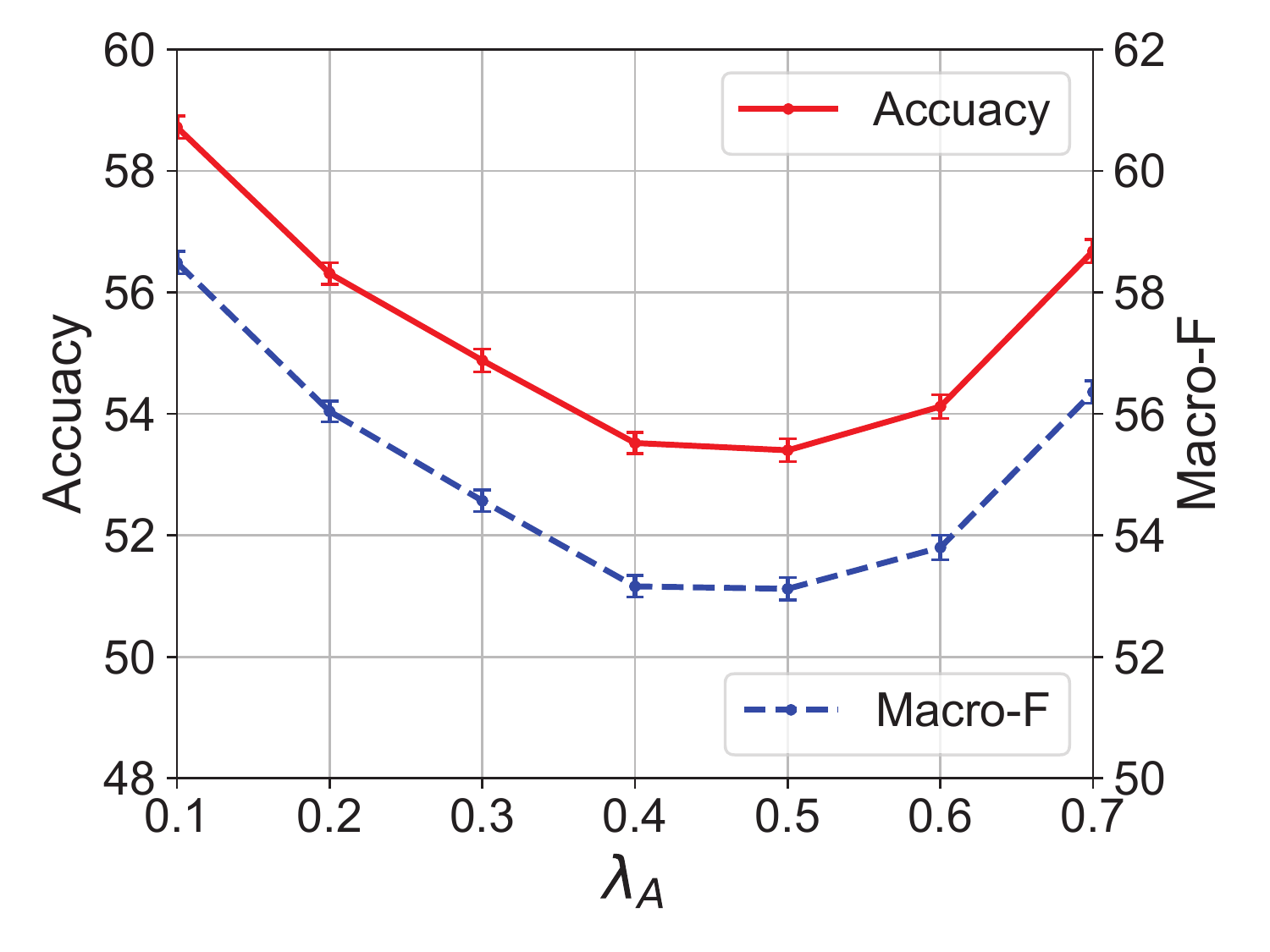}\label{a1}
    }
        \subfigure[Performance.]{
    \includegraphics[width=0.46\linewidth]{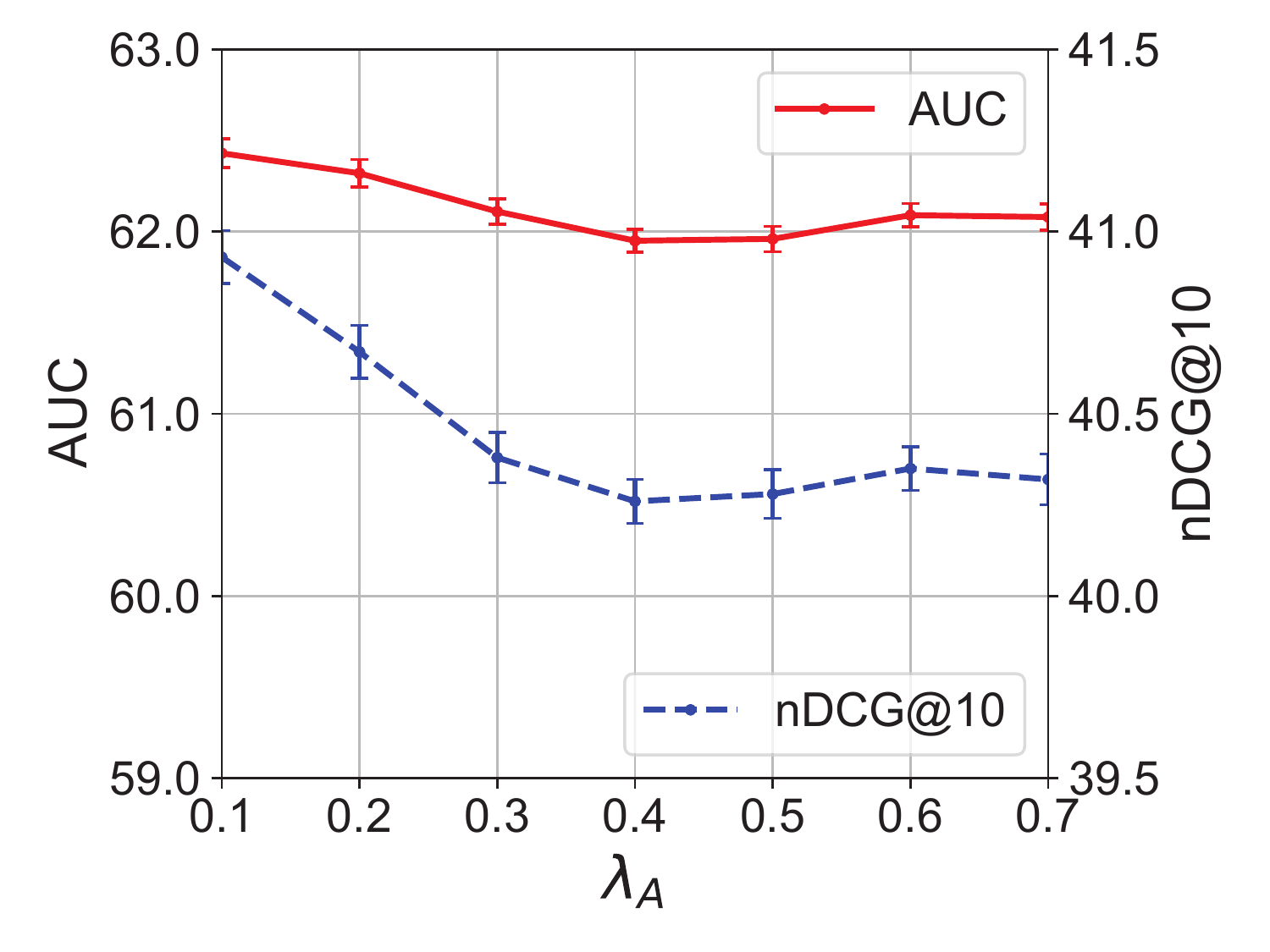}\label{a2}
    }
    \caption{The news recommendation fairness and performance w.r.t. different $\lambda_A$.}
    \label{fig:lambdag}\vspace{-0.1in}
\end{figure}

In this section, we explore the influence of several critical hyperparameters, i.e., the loss coefficients $\lambda_G$, $\lambda_D$ and $\lambda_A$ in Eq.~(\ref{eq1}) on the fairness and performance of news recommendation.
Since there are three hyperparameters, their influence is evaluated independently.
Firstly, we vary the value of $\lambda_G$ without the decomposition loss and adversarial learning, and plot the fairness results under $K=10$ in Figs.~\ref{g1} and~\ref{g2}.
We see the attribute prediction task can help improve the recommendation fairness, and the improvement increases when $\lambda_G$ grows from 0.
However, the improvement is marginal when it is larger than 0.5, and the performance declines more rapidly.
Thus,  a moderate value for $\lambda_G$ (e.g., 0.5) may be preferable to achieve good fairness without too heavy performance loss.
Then, we vary the value of $\lambda_D$ under $\lambda_G=0.5$ and adversarial learning deactivated.
The results are shown in Figs.~\ref{d1} and~\ref{d2}.
From these results, we also find that the recommendation fairness improves with the increasing of $\lambda_D$, and the performance may decline when $\lambda_D$ is too large.
Thus, a proper range of $\lambda_D$ (0.3-0.6) can achieve a good tradeoff between recommendation fairness and performance.
For convenience, we choose the same value for $\lambda_D$ as $\lambda_G$, i.e., 0.5.
Finally, we activate the adversarial discriminator and vary $\lambda_A$ under $\lambda_G=\lambda_D=0.5$.
The results are shown in Figs.~\ref{a1} and~\ref{a2}.
We find that if $\lambda_A$ is too small or too large, the recommendation results are less fair.
This may be because the adversaries cannot achieve an appropriate equilibrium and the attribute label is leaked to the bias-free user embedding.
Thus, a moderate value of $\lambda_A$ is also necessary, and for convenience of hyperparameter selection, we choose $\lambda_A=\lambda_G=\lambda_D=0.5$ to avoid too heavy effort on hyperparameter searching.

\subsection{Case Study}

\begin{figure}[!t]
    \centering
    \includegraphics[width=1.0\linewidth]{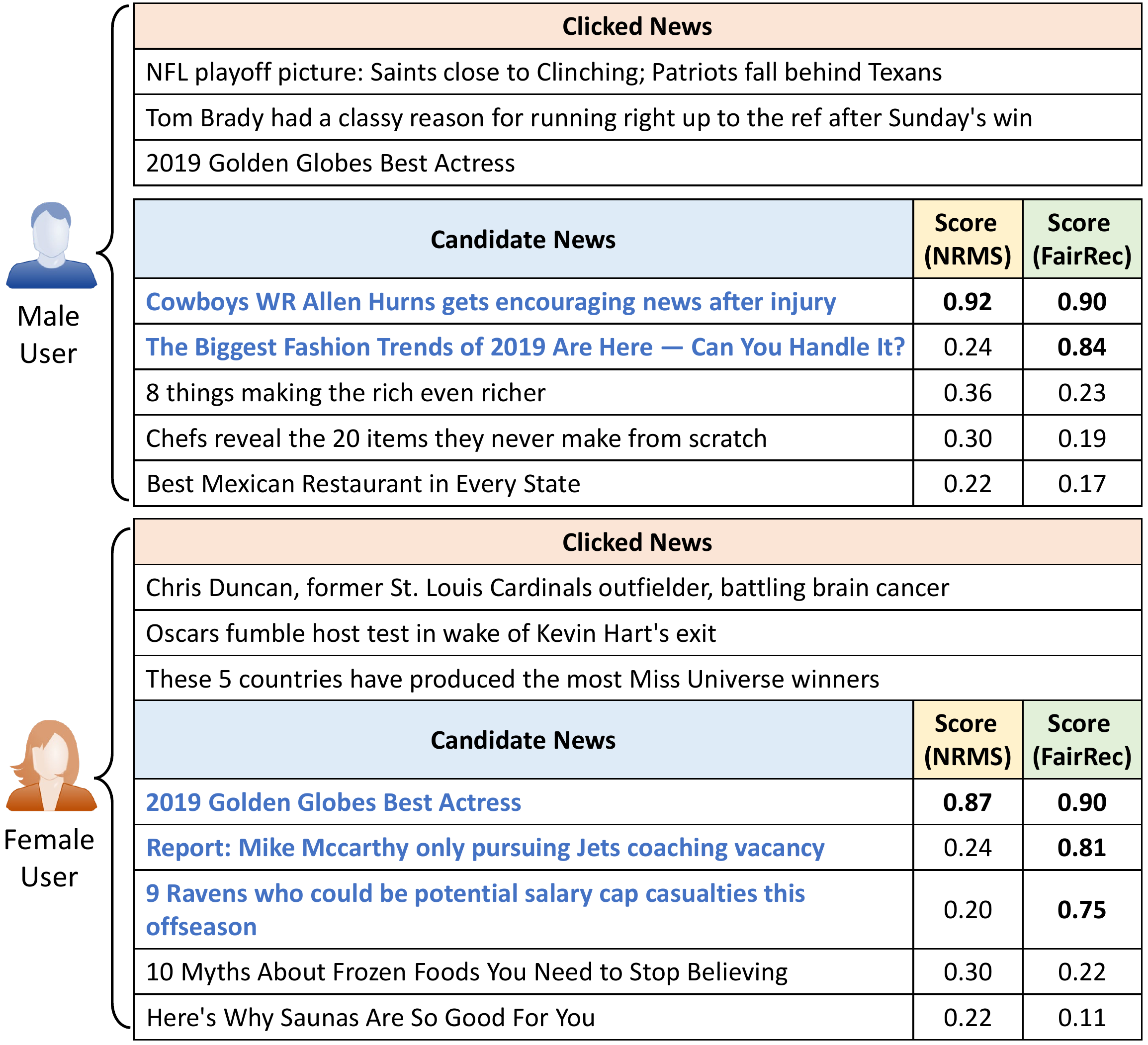}
    \caption{Comparison between the recommendation results of \textit{NRMS} and \textit{FairRec} for a male and a female user. The clicked candidate news are in blue.}
    \label{fig:case}\vspace{-0.1in}
\end{figure}

We conduct several case studies to show that our approach can improve the fairness of news recommendation results.
We randomly select a male user and a female user, and predict the ranking scores of candidate news based on their clicked news using \textit{NRMS} and \textit{FairRec}.
The results are illustrated in Fig.~\ref{fig:case}.
From the top table in  Fig.~\ref{fig:case}, we can infer that this male user may be interested in football and Golden Globes.
However, the NRMS method that does not consider recommendation fairness provides a top rank for the candidate news about sports (Cowboys WR...) while assigns candidate news about fashion (The Biggest...) a low rank, which may be because fashion news is more likely to be preferred by female users.
However, this user may also be interested in this news because it in fact has some inherent relatedness with the clicked news ``2019 Golden Globes Best Actress''.
Similar phenomenon also exists in the ranking results of the female user.
We can infer that this user may be interested in baseball games, and she may also have some interests in football.
However, the news about football is assigned relatively low ranks, since football news may be preferred more by male users.
These results reflect the unfairness in news recommendation.
Fortunately, Fig.~\ref{fig:case} shows that our approach can recommend the fashion news to male users and NFL news to female users for better satisfying their interest.
It indicates that our approach can effectively improve fairness in news recommendation.

\section{Conclusion}

In this paper, we propose a fairness-aware news recommendation approach with decomposed adversarial learning and orthogonality regularization.
We propose to decompose the user interest model into two parallel ones to respectively learn a bias-aware user embedding that captures bias information and a bias-free user embedding for fairness-aware news ranking.
In addition, we apply an attribute prediction task to the bias-aware user embedding to enhance its ability on bias modeling, and apply adversarial learning techniques to the bias-free user embedding to eliminate its bias information on user attributes.
Besides, we propose an orthogonality regularization method that pushes both user embeddings to be orthogonal to each other, which can better remove user attribute information from the bias-free user embedding. 
Extensive experiments show that our approach can substantially improve news recommendation fairness with minor performance sacrifice. 

\section*{Acknowledgments}
This work was supported by the National Natural Science Foundation of China under Grant numbers U1936208, U1936216 and 61862002.

\bibliography{main}
\end{document}